\documentclass[aps,prl,twocolumn,groupedaddress]{revtex4}
\usepackage{graphicx}
\begin{document}

\title{The Vogel-Fulcher-Tammann law in the elastic theory of glass transition}
\author{Kostya Trachenko}
\address{Department of Earth Sciences, University of Cambridge,
Downing Street, Cambridge, CB2~3EQ, UK}

\begin{abstract}
We propose that the origin of the Vogel-Fulcher-Tammann law is the increase of the range of elastic interaction between local relaxation events in a liquid. In this picture, we discuss the origin of cooperativity of relaxation, the absence of divergence of relaxation time at a finite temperature and the crossover to a more Arrhenius behaviour at low temperature.
\end{abstract}


\maketitle

The transition of a liquid into a glass on lowering the temperature may appear conceptually simple, yet this phenomenon has turned out to be one of the most difficult and controversial problems in condensed matter physics, the problem of the glass transition \cite{langer,dyre}. At high temperature, relaxation time $\tau$ of a liquid follows Arrhenius dependence. On lowering the temperature, $\tau$ almost universally deviates from Arrhenius dependence, and follows the Vogel-Fulcher-Tammann (VFT) law:

\begin{equation}
\tau=\tau_0\exp\left(\frac{A}{T-T_0}\right)
\end{equation}

\noindent where $A$ and $T_0$ are constants. The origin of the VFT law is the main open question in the field of the glass transition \cite{langer,dyre}.

A related open question follows from the form of the VFT law, namely what happens at $T_0$. Because $\tau$ formally diverges at $T_0$, several models have suggested that a phase transition from a liquid to a glass phase can exist \cite{langer,dyre}. Because the divergence is not observed in an experiment, it was proposed that the phase transition is avoided due to sluggish dynamics when $\tau$ exceeds experimental time scale. However, the nature of the phase transition and the second phase is not clear, which continues to fuel the current debate \cite{langer,dyre}. Interestingly, the VFT law changes to a more Arrhenius form at low temperature, pushing the divergence temperature down \cite{sti}. The origin of this crossover is not understood.

Another related problem is the physical origin of ``cooperativity''. The notion of cooperativity of molecular motion, which sets in a liquid as temperature is lowered, was introduced and intensely discussed in several popular theories of the glass transition. These theories are based on the assumption that ``cooperatively rearranging regions'', ``domains'' or ``clusters'' exist in a liquid, in which atoms move in some concerted way that distinguishes these regions from their surroundings \cite{langer,dyre,adam,ngai,yama,argon}. The physical origin of cooperativity is not understood, nor is the nature of concerted motion.

A glass is different from a liquid by virtue of its ability to support shear stress. This suggests that the change of stress relaxation mechanism in a liquid on lowering the temperature is central to the glass transition process, yet stress relaxation is not discussed in popular glass transition theories, including entropy, free-volume, energy landscape and other approaches \cite{dyre}.

In this paper, we discuss how stress relaxation in a liquid changes with temperature. We propose that the origin of the VFT law is the increase of the range of elastic interaction between local relaxation events. In this theory, we also discuss the origin of cooperativity of relaxation, the absence of divergence of $\tau$ at a finite temperature and the crossover to a more Arrhenius behaviour at low temperature.

Relaxation and flow in a liquid proceed by elementary local structural rearrangements, during which atoms jump out of their cages. We call these rearrangements local relaxation events (LREs). Because the divergence of the elastic field due to a LRE is zero, a LRE is not accompanied by compression of the surrounding liquid, and can be viewed, in a simple model, as a pure shear event \cite{dyre}. Therefore, in discussing how LREs interact elastically, we consider shear LREs. A typical shear relaxation event is shown in Figure 1 (term ``concordant'' in the figure caption is not important here, and will be explained later). The accompanied structural rearrangement produces elastic shear stress which propagates through the system and affects the relaxation of other events. The important question here is how does this stress affect relaxation of other LREs in the liquid?

\begin{figure}
\begin{center}
{\scalebox{0.55}{\includegraphics{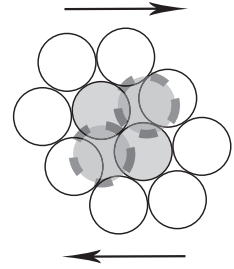}}}
\end{center}
\caption{An example of a concordant local relaxation event. Solid and dashed lines around the shaded atoms correspond to
initial and final positions of a rearrangement, respectively. Arrows show the direction of external stress.}
\end{figure}

Lets consider how the changes of stresses due to remote shear LREs affect a given local relaxing region, shown in the centre in Figure 2. Relaxation of the central event involves deformation of the ``cage'' around the jumping atom (see Figure 1), and therefore depends on the stresses that propagate from the remote LREs to the centre. A remote shear LRE, similar to the one shown in Figure 1, creates elastic shear waves, which include waves of high frequency. This is because the deformation, associated with a LRE, creates a wave with a length comparable to interatomic separations (see Figure 1), and hence with a frequency on the order of the Debye frequency. At high frequency $\omega>1/\tau$, a liquid supports propagating shear waves \cite{frenkel}, which propagate stress and its variations from remote LREs to the central point. If $\tau$ is macroscopically defined as the time of decay of shear stress in a liquid \cite{frenkel,elast}, $d_{\rm el}=c\tau$ gives the length of this decay, where $c$ is the speed of sound. Here, $d_{\rm el}$ gives an estimation of the maximal range over which shear stress decays in a liquid. At the microscopic level, the relevance of $d_{\rm el}=c\tau$ is as follows. A high-frequency shear wave originating from a LRE propagates stress until a remote LRE takes place at the front of the wave, at which point the wave front is absorbed by the remote LRE. Suppose this happens at distance $d_{\rm el}$ from the original LRE. $d_{\rm el}$ can be calculated from the condition of equality of the wave travel time, $d_{\rm el}/c$, and the time at which the remote LRE takes place at point $d_{\rm el}$. The latter time is given by $\tau$, because microscopically, $\tau$ is defined as the average time between two consecutive LREs at one point in space \cite{frenkel}, and we obtain $d_{\rm el}=c\tau$ as before.


Therefore, $d_{\rm el}$ defines the maximal distance over which the central LRE is affected by elastic shear stresses due to other LREs in a liquid (see Figure 2). For this reason, $d_{\rm el}$ can be called the {\it liquid elasticity length}. Note that relaxation of the central event is affected by all those stresses that have enough time to propagate to the centre. Because it takes time $\tau$ for the central event to relax, its relaxation is affected by the stresses from all LREs located distance $c\tau$ away. After time $\tau$, the central event relaxes, and the process repeats. Therefore, the definition $d_{\rm el}=c\tau$ is self-consistent.

\begin{figure}
\begin{center}
{\scalebox{0.45}{\includegraphics{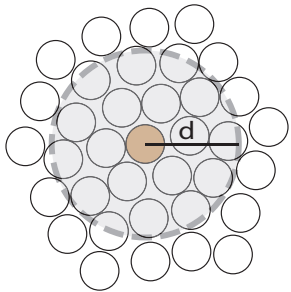}}}
\end{center}
\caption{Illustration of the elastic interaction between local relaxation events. This interaction takes place within the range
$d_{\rm el}$ from the central relaxing regions. Shaded and open circles represent local relaxing regions inside and outside, respectively, of the interaction sphere.}
\end{figure}

Because $c$ is on the order of $a/\tau_0$, where $a$ is the interatomic separation of about 1 \AA\ and $\tau_0$ the oscillation period, or inverse of Debye frequency ($\tau_0\approx 0.1$ ps),

\begin{equation}
d_{\rm el}=c\tau=a\frac{\tau}{\tau_0}
\end{equation}

On lowering the temperature, $\tau$ increases as $\tau=\tau_0\exp(V/kT)$, where $V$ is the activation barrier of a LRE \cite{frenkel} (here, $V$ can be temperature-dependent). According to Eq. (2), this increases $d_{\rm el}$ and the number of LREs that elastically interact with a given event. We propose that this is the key to the super-Arrhenius relaxation.

Before discussing the VFT law itself, we note that Eq. (2) immediately gives the crossover from non-cooperative to cooperative relaxation. When, at high temperature, $\tau\approx\tau_0$, $d_{\rm el}\approx a$ (see Eq. (2)), and $d_{\rm el}<d_m$, where $d_m$ is the distance between neighbouring LREs of about 10 \AA\ ($d_m$ is the distance between the centres of neighbouring molecular cages). This means that LREs do not elastically interact. As $\tau$ increases on lowering the temperature, $d_{\rm el}\ge d_m$ becomes true. At this point, LREs are no longer independent, because relaxation of a LRE is affected by elastic stresses from other events. This discussion, therefore, clarifies the physical origin of cooperativity. Here, we do not need to assume or postulate cooperativity of relaxation as in the previous work \cite{langer,dyre,adam,ngai,yama,argon}. In this picture, relaxation is ``cooperative'' in the general sense that LREs are not independent, but the origin of this cooperativity is the usual elastic interaction. We have recently shown how this interaction gives rise to stretched-exponential relaxation (SER), a universal feature of supercooled liquids \cite{ser}. The crossover from exponential relaxation to SER takes place when $d_{\rm el}=d_m$. According to Eq. (2), $\tau$ at the crossover, $\tau_c$, is a universal value: $\tau_c=\tau_0 d_m/a$. This gives $\tau_c$ of about 1 ps, consistent with the numerous experiments \cite{cross1,casa}.

In order to derive the VFT law, we recall the previous discussion that $V$ is given by the elastic shear energy of a liquid around a LRE \cite{dyre,nemilov,dyre1}. The energy needed for an atom to escape its cage at the constant volume is very large because of the strong short-range interatomic repulsions, hence it is more energetically favourable for the cage to expand, reducing the energy needed for escape. Such an expansion elastically deforms the surrounding liquid, hence $V$ is given by the work of the elastic force needed to deform the liquid around a LRE. Because this deformation does not result in the compression of the surrounding liquid (for the displacement field $\bf u$ created by an expanding sphere, div$(\bf u)=0$), $V$ is given by the background shear energy of the liquid. This was confirmed by the experiments showing that $V$ increases with the liquid shear energy \cite{dyre1}.

We now recall the previous discussion of how LREs redistribute external stress. In discussing creep, Orowan introduced ``condordant'' LREs \cite{orowan}. A concordant shear LRE is accompanied by a strain in the direction agreeing with the applied external stress, and reduces the local stress and energy (see Figure 1). In order to counter-balance this decrease, other local regions in a system support more stress \cite{orowan}. Goldstein applied the same argument to a viscous liquid under external stress \cite{gold}. Consider that this stress is counterbalanced by stresses supported by local regions. Because a local region supports less stress after a concordant LRE than before, other local regions in the liquid should support more stress after that event than before in order to counter-balance \cite{gold}.

Lets consider a liquid perturbed by a pulse of an external field. At time zero, shear stresses supported by local regions counterbalance external shear stress. As relaxation proceeds, each concordant shear LRE reduces stress locally, until the external stress is relaxed by a certain number of LREs $N$. When this process is complete, the liquid relaxes to equilibrium. At times smaller than $L/c$, where $L$ is the system size, the external stress can be considered constant, and the stress redistribution argument of Orowan-Goldstein applies. Alternatively, we can consider an external stress constantly compensating for the decreases of local stresses. In the resulting steady flow, $\tau$ is the time needed to relax an increment of external perturbation, and can be viewed as the time of the liquid's retardation behind the external field. Let $n$ be the current number of LREs, such that $n\rightarrow N$. If $\Delta p$ is the increase of shear stress on the liquid around a current local region that arises from the need to counter-balance the decreases of stresses due to previous remote concordant LREs, $\Delta p$ increases with $n$. The increase of $\Delta p$ consistently increases the elastic strain in the direction of external shear stress, increasing the background shear energy of the liquid around the current local region. As discussed above, $V$ for the current LRE increases as a result. The increase of $V$, $\Delta V$, due to $\Delta p$ is the work $\int \Delta p {\rm d}q$. If $q_a$ is the activation volume \cite{dyre1}, $\Delta V=\Delta p q_a$, and $V=V_0+q_a\Delta p$, where $V_0$ is the high-temperature activation barrier. Because $\Delta p$ increases with $n$, $V$ also increases with $n$. This gives the {\it elastic feed-forward interaction mechanism} for LREs, which sets SER \cite{ser}.

To calculate $V$ as a function of $d_{\rm el}$, lets consider the last LRE that relaxes an increment of external shear stress to be in the centre of a sphere of radius $d_{\rm el}$ (see Figure 2). As relaxation proceeds, the shear stress on the central region increases in order to counterbalance stress decreases due to previous remote concordant LREs. Importantly, because this mechanism operates in the range set by $d_{\rm el}$ and because $d_{\rm el}$ increases on lowering the temperature (see Eq. (2)), stresses from an increasing number of remote LREs need to be counterbalanced by the central region. It is also important to note that all stresses within a distance $d_{\rm el}=c\tau$ have enough time to propagate to the centre and affect relaxation of the central event (recall self-consistency in definition of $d_{\rm el}$).

Let $\Delta p_i(0)$ be the reduction of local stress due to a remote concordant LRE $i$. $\Delta p_i$ decays with distance, hence we denote $\Delta p_i(r)$ as its value at the centre in Figure 2. The increase of stress on the central rearranging region, $\Delta p$, can be calculated as

\begin{equation}
\Delta p=\rho\int\limits_{d_0/2}^{d_{\rm el}} 4\pi r^2 \Delta p_i (r) {\rm d}r
\end{equation}

\noindent where $\rho$ is the density of local rearranging regions and $d_0$ is on the order of the size of a relaxing region
(in Figure 1, $d_0\ge 3a$). Note that in Eq. (3), $d_{\rm el}$ is the upper integration limit. In what follows, we assume, for simplicity, that $\Delta p_i(0)$ are constant, $\Delta p_i(0)=\Delta p_0$.

In an elastic medium, stresses decay as $\Delta p (r)\propto 1/r^3$ \cite{elast}. Because $\Delta p(r)=\Delta p_0$ at $d_0/2$,
$\Delta p(r)=\Delta p_0(d_0/2r)^3$. Integration of Eq. (3), together with $V=V_0+q_a\Delta p$ from the discussion above, gives

\begin{equation}
V=V_0+\pi/2\rho q_a\Delta p_0 d_0^3 \ln(2d_{\rm el}/d_0)
\end{equation}

Using $\tau=\tau_0\exp(V/kT)$ in Eq. (2), we obtain

\begin{equation}
d_{\rm el}=a\exp\left(\frac{V}{kT}\right)
\end{equation}

Eqs. (4) and (5) define $V$ in a {\it self-consistent} way. Eliminating $d_{\rm el}$ from the two equations, we find:

\begin{equation}
V=\frac{AT}{T-T_0}
\end{equation}

\noindent where $A=V_0+\pi/2\rho q_a\Delta p_0 d_0^3\ln(2a/d_0)$ and $kT_0=\pi/2\rho q_a\Delta p_0 d_0^3$.

From Eq. (6), the VFT law follows:

\begin{equation}
\tau=\tau_0\exp\left(\frac{A}{T-T_0}\right)
\end{equation}

In this picture, the super-Arrhenius behaviour is related to the increase of $d_{\rm el}$ (see Eq. (4)). The transition from the VFT law to the Arrhenius form of $\tau$ takes place in the limit of small $d_{\rm el}$ at high temperature. In this case, the upper and lower integration limits in Eq. (3) coincide, giving $\Delta p=0$, $V=V_0$ and $\tau=\tau_0\exp(V_0/kT)$.

In the proposed theory of the glass transition, the ongoing controversy \cite{langer,dyre,angell} regarding the divergence and possible phase transition at $T_0$ is readily reconciled. The divergence at $T_0$ can not exist for the following reason. From Eqs. (5,6), we find

\begin{equation}
d_{\rm el}=a\exp\left(\frac{A}{T-T_0}\right)
\end{equation}
\noindent

\noindent When $T$ approaches $T_0$, $d_{\rm el}$ diverges, and quickly exceeds any finite size of the system $L$. When $d_{\rm el}\ge L$, all LREs in the system elastically interact, and there is no room for the increase of $V$ by way of increasing $d_{\rm el}$. The upper limit of integral (3) becomes $d_{\rm el}=L$, giving temperature-independent $V\propto \ln(L)$ (see Eq. (4)). Further decrease of temperature has a weaker effect on $V$, and can be due to, e.g., density increase, but not to the increase of $d_{\rm el}$ (the density-related contribution to $V$ does not depend on $d_{\rm el}$ or $L$). As a result, the behaviour of $\tau$ tends to Arrhenius, pushing the divergence to zero temperature.

$d_{\rm el}$ exceeds the experimental value of $L$ above $T_g$: if $\tau(T_g)=10^3$ sec, $d_{\rm el}(T_g)=10^3$ km, according to Eq. (2). Hence our theory predicts the crossover from the VFT law to a more Arrhenius behaviour at low temperature, as is seen in the experiments \cite{sti}. According to Eq. (2), $\tau$ at the crossover is $\tau=\tau_0 L/a$. If a typical value of $L$ is 1 mm, $\tau$ at the crossover is $10^{-6}$ sec, consistent with the experimental results \cite{cross2}.

We note here that $d_{\rm el}$ vastly exceeds the size of ``cooperatively rearranging regions'' (CRR), which is several nm at $T_g$ (for review, see, e.g., Ref. \cite{yama}). The physical picture of CRR is not clear \cite{dyre}. It is possible that the observed nm scale of CRR is set by the distance beyond which the elastic strains from LREs decay to the values undistinguishable from thermal fluctuations.

$d_{\rm el}$ gives an insight into the origin of liquid fragility \cite{angell}. According to Eq. (4), as long as at high temperature $d_{\rm el}<L$, lowering the temperature increases $V$, resulting in a fragile behaviour. If, on the other hand, $d_{\rm el}\ge L$ at high temperature already, further decrease of temperature has a weaker effect on $V$, giving weak super-Arrhenius behaviour. Experimentally, for many systems the studied range of temperatures varies from about $2T_g$ and $T_g$ \cite{casa}, hence we consider the increase of $d_{\rm el}$ from high temperature $T_h=2T_g$ to $T_g$. Take, for example, two systems on the stronger side of fragility plots, BeF$_2$ and SiO$_2$. From the experimental values of $V_h/kT_g$ ($V_h$ is the activation barrier at the highest measured temperature), we find $V_h/kT_h=24$ and 19.6 for BeF$_2$ and SiO$_2$, respectively \cite{novikov}. According to Eq. (5), this gives $d_{\rm el}=2.6$ m and 33 mm at $T_h$ for the two systems. Because a typical experimental value of $L$ is on the order of 1 mm, our theory correctly predicts that these systems should be on the strong end of fragility plots. For two fragile systems, toluene and propylene carbonate, $V_h/kT_h=3.34$ and 5.75, giving $d_{\rm el}=28$ and 314 \AA\ at $T_h$, respectively. This is much smaller than $L$, hence our theory predicts that these systems should be fragile, as is seen experimentally. An interesting prediction from this picture is that strong systems will show increased fragility at high temperature when $d_{\rm el}<L$ (note that strong systems have been measured at relatively low temperature only \cite{angell}).

Before concluding, we note that we discussed a typical experimental setup, in which a liquid is perturbed and $\tau$ is measured as the time of return to equilibrium. All above results remain the same in the equilibrium case as well, when thermally fluctuating LREs interact via induced elastic stresses in the range set by $d_{\rm el}$ \cite{future}.

In summary, we proposed that the origin of the Vogel-Fulcher-Tammann law is the increase of the range of elastic interaction between local relaxation events in a liquid. In this picture, we discussed the origin of cooperativity of relaxation, the absence of divergence of relaxation time at a finite temperature and the crossover to an Arrhenius behaviour at low temperature.

We suggest that the proposed theory is applicable to other systems, in which local rearrangements interact via the fields they induce. This includes a wide range of phenomena, for example, relaxation in spin glasses. Here, the same universal relaxation effects as in structural glasses are observed, including the VFT law, cooperativity, SER and other phenomena.

I am grateful to V. V. Brazhkin, A. Kehagia, R. Casalini and C. M. Roland for discussions, and to EPSRC for support.

\end{document}